\documentclass[12pt,preprint]{aastex}
\begin{document}

\title{Six New Galactic Orbits of Globular Clusters in a
 Milky-Way-Like Galaxy}

\author{Christine Allen\altaffilmark{1}, Edmundo
Moreno\altaffilmark{1}, and B\'arbara Pichardo\altaffilmark{1,2} }
\altaffiltext{1}{Instituto de Astronom\'\i a, Universidad Nacional
Aut\'onoma de M\'exico, M\'exico, D.~F.,
M\'exico.}\altaffiltext{2}{Institute for Theoretical Physics,
University of Zurich, Winterthurerstrasse 190, Zurich 8057,
Switzerland.}

\begin{abstract}

  Absolute proper motions for six new globular clusters have recently
  been determined.  This motivated us to obtain the Galactic orbits of
  these six clusters both in an axisymmetric Galactic potential and in
  a barred potential, such as the one of our Galaxy. Orbits are also
  obtained for a Galactic potential that includes spiral arms.  The
  orbital characteristics are compared and discussed for these three
  cases. Tidal radii and destruction rates are also computed and
  discussed.

\end{abstract}


\section{INTRODUCTION}\label{introd}
Absolute proper motions have become available for six new globular
clusters \citep{CD07}, thus increasing to 54 the number of globular
clusters of our Galaxy for which full space velocities and Galactic
orbits can now be calculated. In a previous paper (Allen, Moreno \&
Pichardo 2006, hereafter Paper I) Galactic orbits were computed for 48
globular clusters, using both an axisymmetric and a barred
Milky-Way-like potential.  We found that the effect of the bar was
greatest for clusters with orbits residing mostly within the bar, and
was negligible for the outermost clusters of our sample.

In the interim, \citet{CD07} have obtained absolute proper motions for
six additional clusters; they have also computed Galactic orbits in an
axisymmetric potential. Since most of their cluster orbits appear to
reside quite within the region of influence of the Galactic bar, it
seems worthwhile to compute the orbits in a barred potential. Also, by
computing orbits using different axisymmetric Galactic models some
insight might be gained on the sensitivity of the results upon the
potential model adopted.

In particular, \citet{CD07} comment on the apparent ``pairing'' of the
orbital parameters of NGC 2808 and NGC 4372, on one hand and NGC 4833,
and NGC 5986 on the other. Since these ``pairings'' or kinematic
groups may have important implications for the dynamical and merger
history of our Galaxy (Kepley et al. 2007; Allen et al. 2007, and
references therein), it is interesting to assess whether or not these
similarities are sensitive to the Galactic model potential used.

In the present paper we compute Galactic orbits for the six new
clusters, in the axisymmetric potential of \citet{AS91}, in the barred
model of \citet{PMM04}, as well as in a potential model including
spiral arms \citep{PMME03}.

As before, we find that the orbits in the barred potential generally
do not show secular changes in the total energy, $E$, or in the
$z$-component of the angular momentum, $h$, both computed in the
inertial Galactic frame.

Since the permanence of bar-like structures in galaxies is a matter of
debate, we cannot be sure that the bar of our Milky has existed
throughout the Galactic lifetime.  Therefore, we also ask ourselves
how the orbits would look like if the bar has existed for only about a
third of this lifetime.

To investigate the possible effects of spiral structure on the orbital
characteristics of the clusters, we also include computations of the
six globular clusters orbits in a Galactic potential that incorporates
spiral arms. Among the six clusters selected for this computation,
there is at least one cluster clearly belonging to the thick disk for
which the effect of spiral perturbations is expected to be large (NGC
5927). 

This paper is organized as follows. In Section \ref{gpot} the Galactic
potentials used to compute the cluster orbits are briefly described,
and the initial conditions are presented.  In Section \ref{galorb} the
orbits obtained with both the axisymmetric and the barred potential
are presented and compared with the ones obtained by \citet{CD07}. In
particular, we show that the effect of the bar is only negligible for
one cluster NGC 3201, the most energetic one. We also discuss the
effects on the orbits of a bar that has existed only during a fraction
of the Galactic lifetime. In Section \ref{spiral} the effect of spiral
structure is presented for all six orbits. In Section \ref{tirad}
tidal radii are computed for orbits in the barred potential and
compared with the axisymmetric case.  Destruction rates for these
clusters are also computed and discussed. The final Section
\ref{conclusiones} presents a brief discussion and our conclusions.

\section{THE GALACTIC POTENTIALS AND THE INITIAL
  CONDITIONS}\label{gpot}

For our study we will employ the axisymmetric Galactic potential of
\citet{AS91}, the barred Galactic potential of \citet{PMM04}, and the
bar-plus-spiral-arms model of \citet{PMME03}.

The bar model (for details see Pichardo et al. 2004) includes a bar of
3.13 kpc scale length, with axial ratios of 1.7:0.64:0.44, and a
conservatively estimated mass of $\sim 10^{10} M\odot$ that replaces
70\% of the spherical bulge mass. The bar moves with an angular
velocity of 60 km s$^{-1}$ kpc$^{-1}$. This bar closely approximates
Model S of \citet{F98}. Again, we use the superposition model of
Pichardo et al. (2004).

To model the spiral perturbation we proceed as \citet{PMME03}, who
refined their model until orbital self-consistent solutions were
found. The spiral arms are constructed by using a three-dimensional
mass distribution obtained by superposing oblate spheroidal potentials
as building blocks of the global spiral potential. For an extensive
description of the model see \citet{PMME03}. The adopted parameters
are given in Section \ref{spiral}.

To calculate the initial conditions we take the absolute proper
motions provided by \citet{CD07}. Other relevant data are taken from
the compilation by \citet{H96}. Once the space velocities are
obtained, we integrate the orbits backwards in time for 1.5 $\times
10^{10}$ years. As in Paper I we analyze the time-reversed orbits.  
For the integration we use the Bulirsch-Stoer
algorithm of \citet{P92}. In the axisymmetric case the relative errors
in the total energy were of about $10^{-14}$ at the end of the
integration. In the barred and spiral potential, the orbits are
computed in the non-inertial reference frame where the bar is at rest.
In the barred potential the precision of the calculations can be
checked using Jacobi's constant. The relative errors in this quantity
turn out to be, typically, $10^{-10}$ to $10^{-11}$.

\section{THE GALACTIC ORBITS}\label{galorb}

\begin{figure}
\plotone{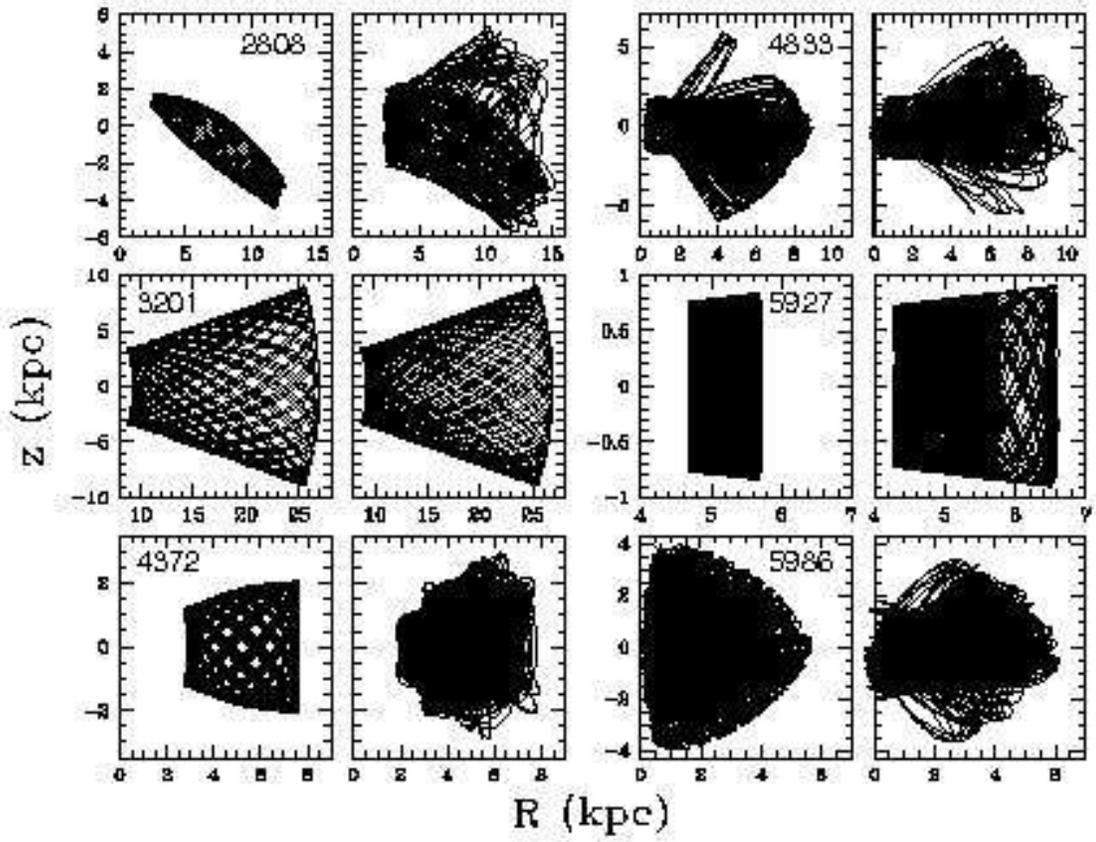}
\caption {Meridional Galactic orbits of the six clusters during the
lifetime of the Galaxy. For each cluster, the orbits in the
axisymmetric (left) and barred (right) potentials are shown.}
\label{mergalorb}
\end{figure}

Figure \ref{mergalorb} shows the meridional orbits of the six clusters
in both the axisymmetric and the barred potentials. Tables \ref{tbl-1}
and \ref{tbl-2} summarize our results for both cases. In Table
\ref{tbl-1}, corresponding to the axisymmetric potential, successive
columns contain the name of the cluster, the minimum perigalactic
distance reached in the course of the complete orbit, the average
perigalactic distance, the maximum apogalactic distance, the average
apogalactic distance, the maximum distance from the Galactic plane
reached throughout the entire orbit, the average maximum distance from
the Galactic plane, the average orbital eccentricity, the orbital
energy per unit mass, the $z$-component of angular momentum per unit
mass, two values for the computed tidal radii (see discussion in
Section \ref{tirad}), and the observed limiting radius, given in
\citet{H96}.

In Tables \ref{tbl-1} and \ref{tbl-2} two additional lines are given
for each cluster. These correspond to ``extreme'' orbits, that take
into account observational uncertainties. The meaning of these
``extreme orbits'' is discussed later in this section.  Table
\ref{tbl-2}, for the barred potential, is similar; but since neither
the orbital energy, $E$, nor the $z$-component of angular momentum,
$h$ (both computed in the inertial Galactic frame) are constants of
motion, we give in columns 9 and 10 the minimum and maximum values
attained by $h$ in the course of the complete orbit.

Only the orbit of NGC 3201, the least bound of the clusters, is not
noticeably affected by the bar. The orbit remains entirely outside the
bar region. This orbit closely resembles that of NGC 4590, presented
in Paper I. As can be seen in Fig. \ref{mergalorb}, the orbits of the
five remaining clusters are significantly affected by the bar. Their
orbital parameters show fairly large changes as compared to the
axisymmetric case, tending to reach higher values of apogalactic
distance and distance to the Galactic plane, as well as becoming
noticeably more irregular. They resemble the class of ``inner''
clusters studied in Paper I.

The orbits of NGC 4833 and NGC 5986 are irregular both without and
with the bar. The orbit of NGC 2808 is near-resonant in the
axisymmetric case, but it seems more irregular in the barred
potential.  The bar also causes the orbit to reach radii and
$z$-distances larger than in the axisymmetric case.  This orbit
resembles that of NGC 6382, presented in our earlier study.

NGC 5927, a thick disk cluster, has a tightly confined box-type orbit
in the axisymmetric case. The bar causes the orbit to extend to larger
radial distances, reaching almost 7 kpc.

Finally, the orbit of NGC 4372, clearly near-resonant in the
axisymmetric case, becomes irregular with the bar, and reaches larger
$z$-distances.

\begin{figure}
\plotone{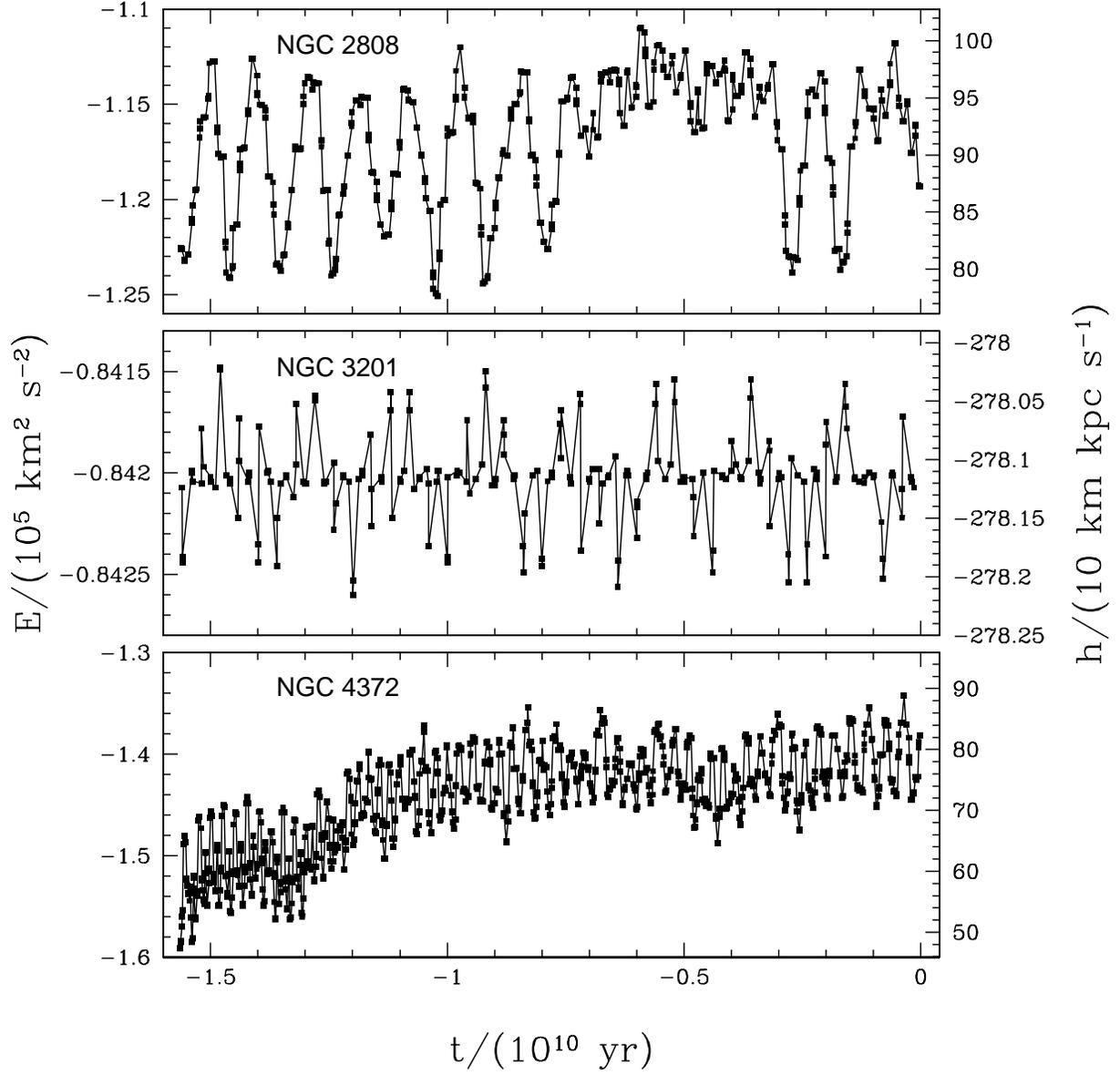}
\caption {Energy and $z$-angular momentum as functions of time in NGC
2808, NGC 3201, and NGC 4372, with the barred potential.}
\label{ehvst_bar}
\end{figure}

\begin{figure}
\plotone{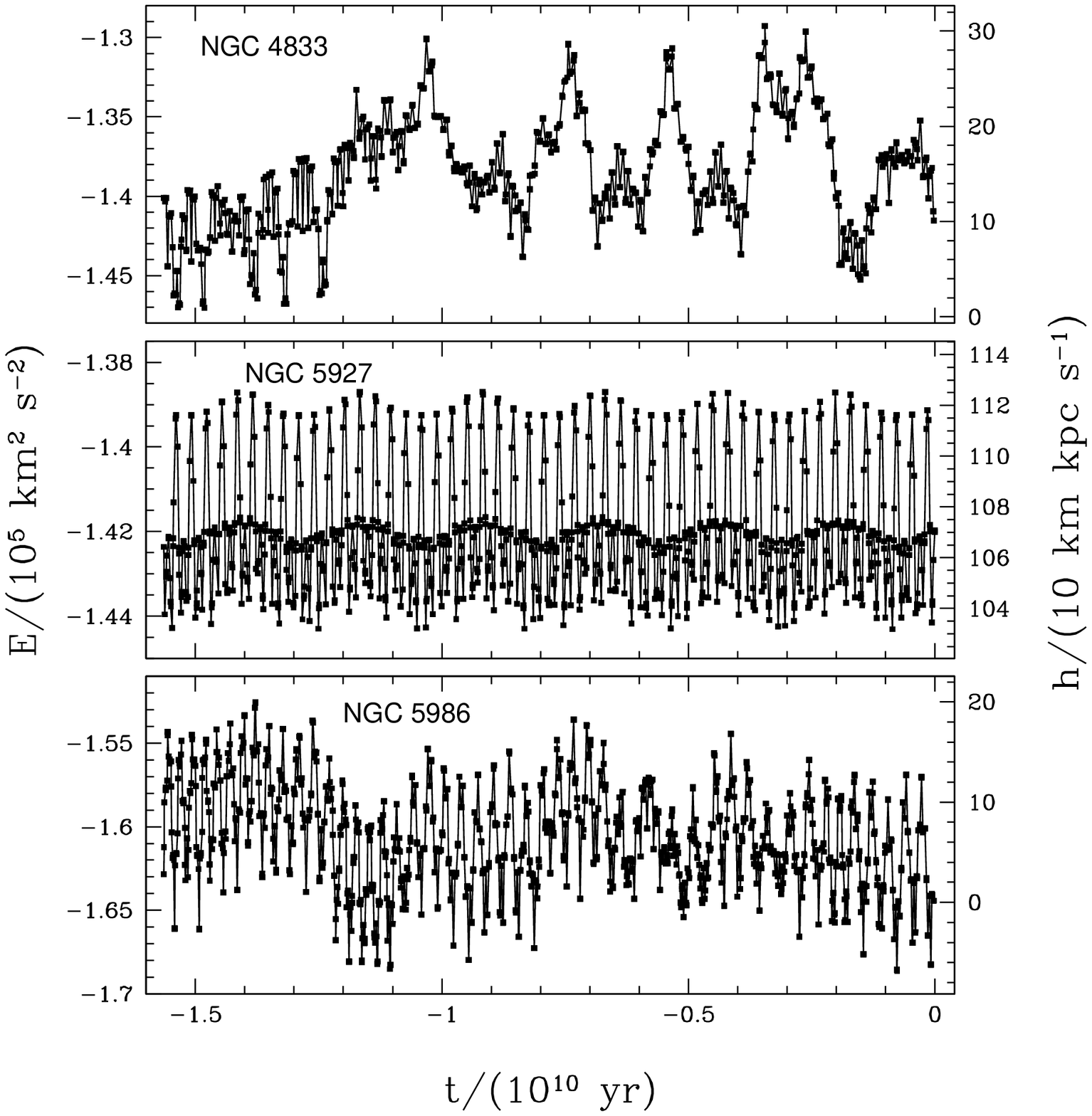}
\caption {As in Figure \ref{ehvst_bar}, for NGC 4833, NGC 5927, and
  NGC 5986.}
\label{ehvst_bar2}
\end{figure}

Plots of the run of energy, $E$, and $z$-component of angular
momentum, $h$, were obtained for all clusters. They are shown in
Figures \ref{ehvst_bar} and \ref{ehvst_bar2}.  These quantities are,
of course, conserved in the axisymmetric case. In the presence of the
bar, the orbits generally do not show large secular changes in $E$ or
$h$. Indeed, these quantities are conserved on the average within
better than 10 percent.  But periodic or quasi-periodic changes are
seen to occur in all cases, reflecting the interactions with the
bar. In Figures \ref{ehvst_bar} and \ref{ehvst_bar2} we can see that
NGC 3201, and NGC 5927 show small periodic or quasi-periodic
changes. The occasional sudden changes we found in our previous study
were found again for four of the six orbits here examined namely for
NGC 2808, NGC 4372, NGC 4833 and NGC 5986. For this last cluster, the
one with the smallest angular momentum, the abrupt changes in the
angular momentum occasionally reverse the sense of revolution of the
cluster around the Galactic center.

\begin{figure}
\plotone{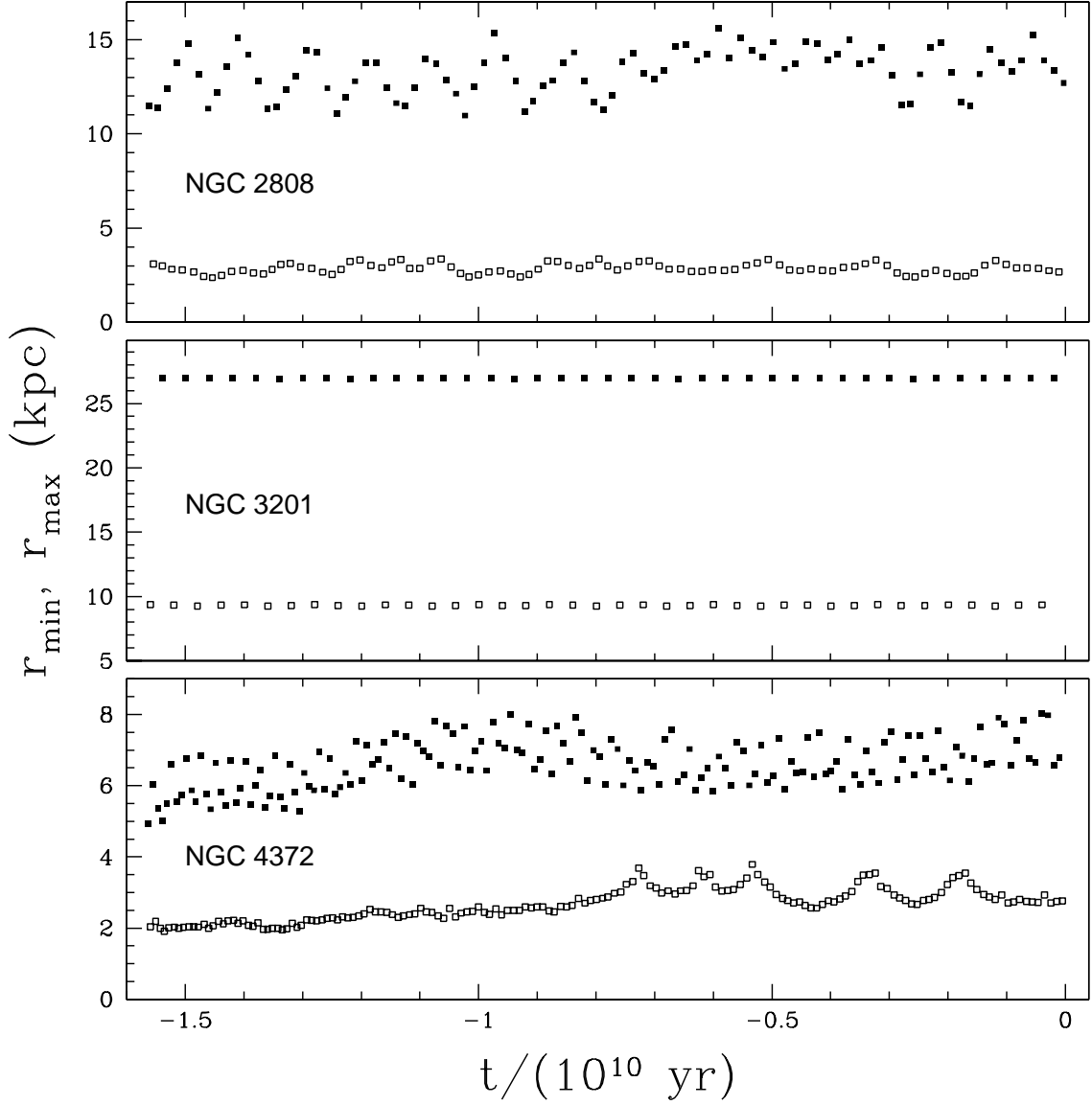}
\caption {Galactocentric distances $r_{min}$ (empty squares) and
$r_{max}$ (filled squares) in NGC 2808, NGC 3201, and NGC 4372.}
\label{rminmax1}
\end{figure}

\begin{figure}
\plotone{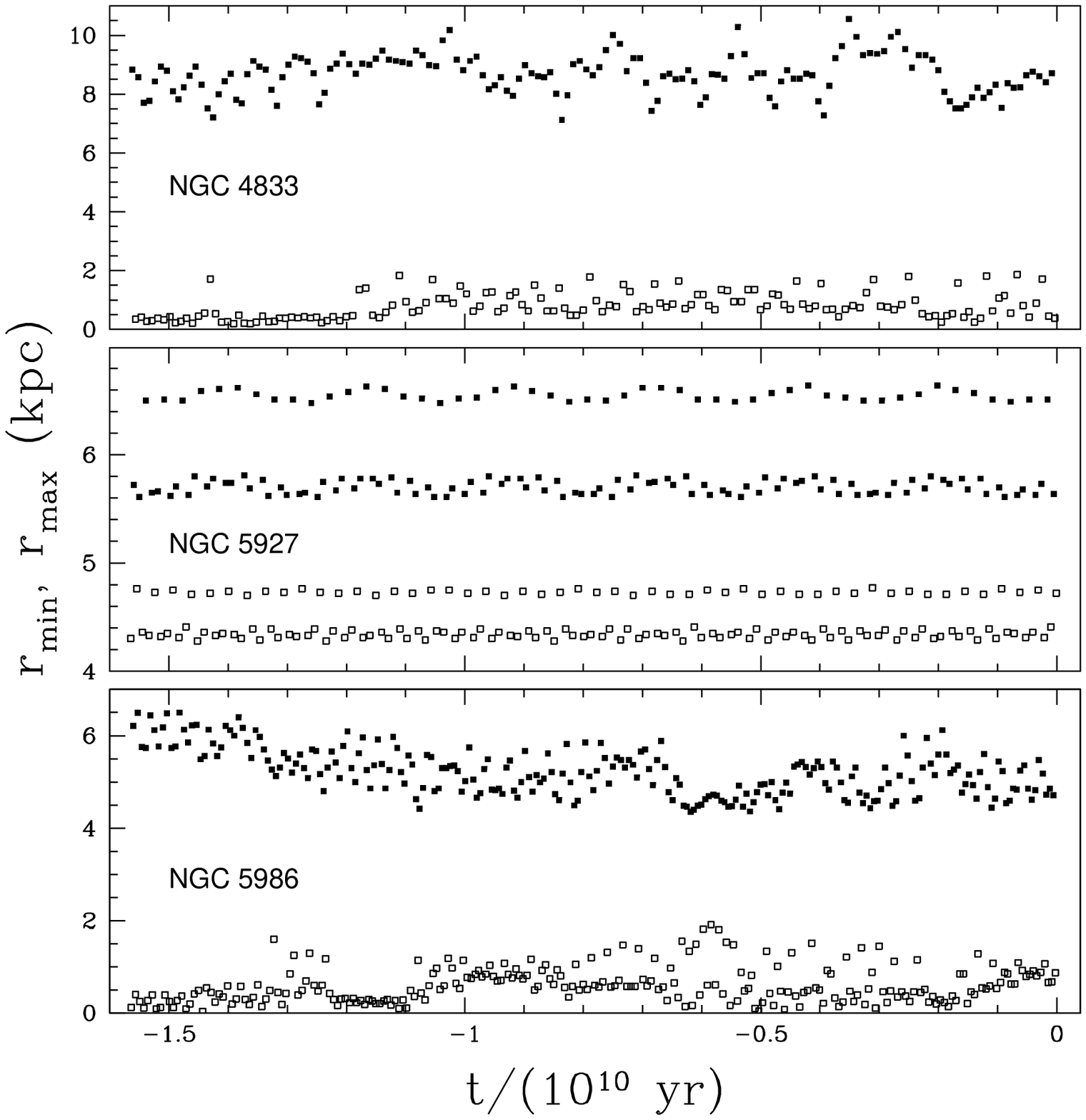}
\caption {As in Figure \ref{rminmax1}, for NGC 4833, NGC 5927, and
NGC 5986.}
\label{rminmax2}
\end{figure}

\begin{figure}
\plotone{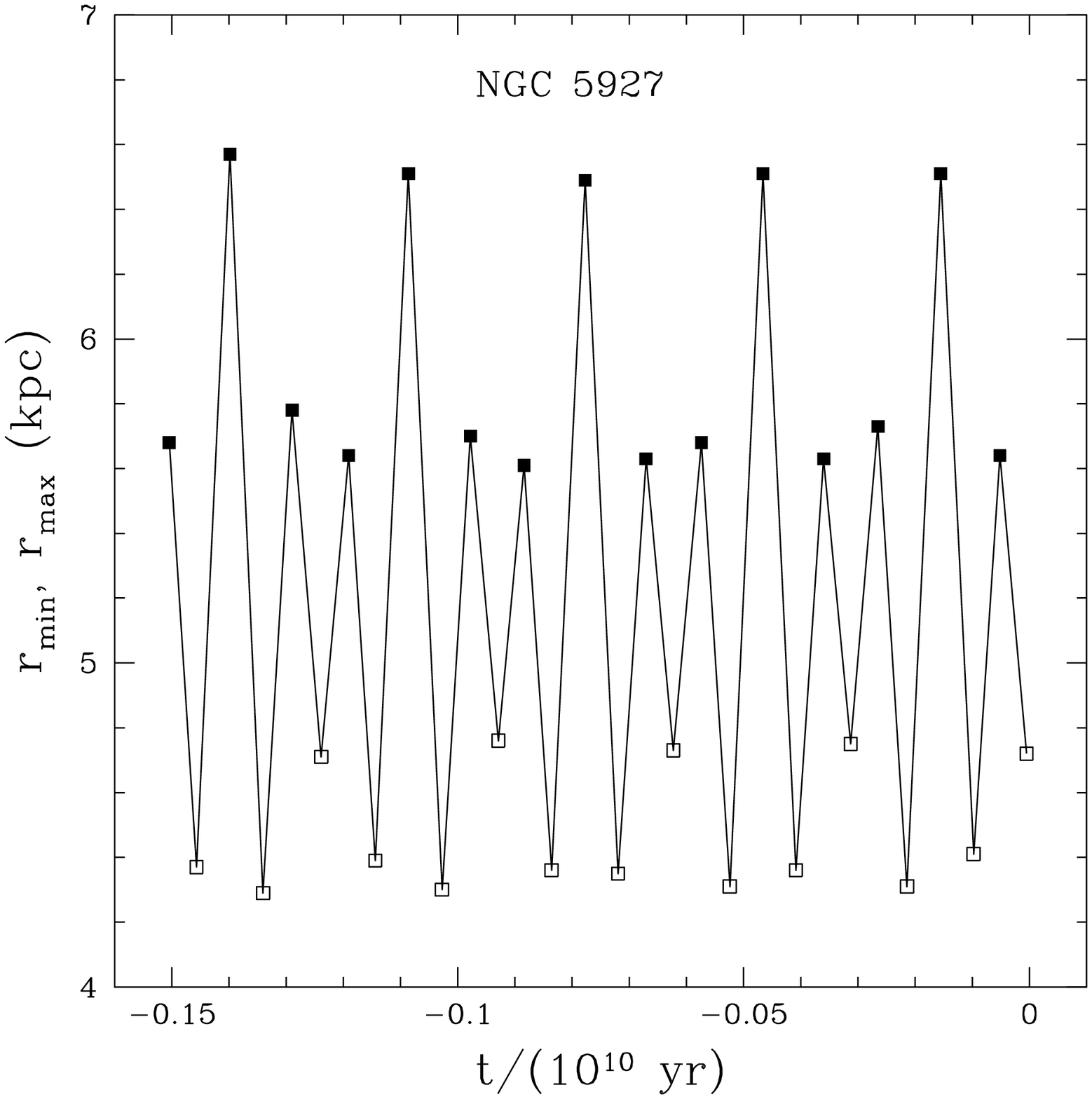}
\caption {Detail around $t = 0$ of the middle frame in Figure
\ref{rminmax2} for NGC 5927. The continuous line connects successive
$r_{min}$ and $r_{max}$.}
\label{rminmax3}
\end{figure}

Figures \ref{rminmax1} and \ref{rminmax2} show plots of $r_{min}$ and
$r_{max}$, the minimum and maximum galactocentric distances attained
by the six clusters during their time-reversed orbits in the barred
potential.  The figures show that these distances oscillate,
particularly $r_{max}$. Clusters showing abrupt energy changes, like
NGC 2808, NGC 4372, NGC 4833 and NGC 5986 also show changes in the
oscillation of $r_{max}$, $r_{min}$, and these occur at about the same
times as the abrupt energy changes.  Figure \ref{rminmax3} shows a
blowup of these oscillations for a short time at the beginning of the
orbit computation of NGC 5927. The figure shows that the cluster
attains the smaller $r_{max}$ twice as often as the larger $r_{max}$.
It reaches the smallest $r_{min}$ twice as often as the larger
$r_{min}$.  This regular behavior is established right from the
beginning of the orbit computation.

\begin{figure}
\plotone{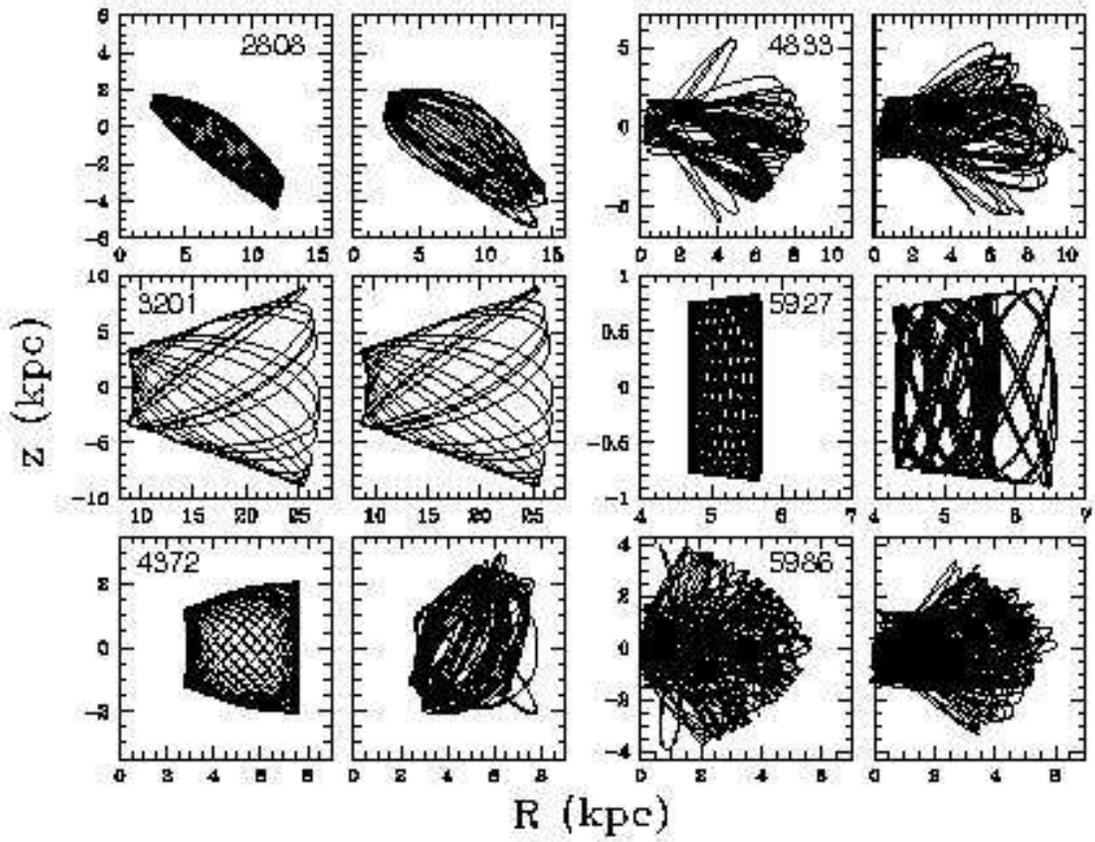}
\caption {As in Figure \ref{mergalorb}, now for the last 5 Gyr.}
\label{mergalorb_sh}
\end{figure}

To obtain a very rough idea of the changes in the orbits that would
appear if the Galactic bar were not a permanent feature, we computed
the orbits backward in time, but only for the last 5.0 Gyr, that is,
approximately a third of the Galactic lifetime.  Figure
\ref{mergalorb_sh} shows the orbits we obtained.  A comparison of
Figure \ref{mergalorb} and Figure \ref{mergalorb_sh} clearly shows
that the effects of the bar are quite similar for both cases. The sole
exception is NGC 2808 whose orbit, when integrated backwards for only
5 Gyr, remains close to a resonance.  We conclude that even if the
Galactic bar has existed only for the last third of the Galactic
lifetime, its effect on the orbits of the ``inner'' clusters is quite
as noticeable as that of a bar present throughout the history of the
Galaxy.

Of course, the bar is not expected to have suddenly come into
existence 5 Gyr ago.  A more realistic simulation would have to take
into account the gradual formation of the bar (and perhaps even its
gradual disappearance and re-formation).  However, this kind of
simulation is well beyond the scope of the present work.  The issue of
the origin and persistence of bars in galaxies is currently a subject
of lively debate.

To assess the effects of observational uncertainties on the orbital
parameters, we computed, as in Paper 1, two additional orbits for each
cluster. The initial conditions for these orbits were chosen so as to
maximize and minimize the orbital energy. In other words, we combined
the observational uncertainties in such a way as to obtain two extreme
orbits. Errors in the orbital parameters resulting from observational
uncertainties are expected to be bounded by these extreme orbits, the
real errors being expected to be much smaller. Indeed, note that the
errors in the galactic parameters estimated by Casetti-Dinescu et al.
(2007) for their orbits are much smaller than the ones that could be
inferred from our extreme orbits.

A comparison of the values given in our Table 1 with the Galactic
parameters obtained by \citet{CD07} shows satisfactory overall
agreement. In general, the uncertainties resulting from observational
errors are larger than the ones resulting from using different
axisymmetric Galactic potential models.  However, our results on the
orbits of the cluster ``pairings'' they find (NGC 5986-NGC 4833 and
NGC 2808-NGC 4372) are not sufficiently similar to make a convincing
case for these ``pairings''. The orbital parameters in the barred
potential differ even more. So, we have to regard the ``pairings'' as
not supported by the orbital parameters we obtain here.

\section{THE EFFECTS OF SPIRAL ARM PERTURBATIONS}\label{spiral}

To study possible effects of spiral structure on the orbital
characteristics of the clusters we computed some of the orbits
using a Galactic potential that includes spiral arms. {\it A priori},
one would expect the effect of spiral structure to be small for two
reasons: (a) the mass of the spiral features is small compared to that
of the disk; (b) the great majority of the globular clusters spend
most of their lifetimes away from the Galactic plane, and hence from
the region of influence of the spiral arms. However, this is not the
case for the thick-disk globular clusters. In this section we
calculate the effect of spiral arms on the orbits of the six new
clusters.

\subsection{The Spiral Model}\label{modelo}

For the orbit computations we have used the semi-analytical model of
Pichardo et al. (2003). This model is based on a three-dimensional
mass distribution composed by the superposition of oblate spheroids as
building blocks of the global spiral potential, rather than on a
simple cosine law as it is customary. It provides a realistic 
representation of spiral features, particularly of flocculent
arms.

To obtain a model matching current observational data we have updated
the parameters used in Pichardo et al. (2003) according to recent
observational work. In Pichardo et al. (2003) information on
parameters not directly given by observations was completed by
performing orbital calculations and searching for the most orbitally
self-consistent model. The adopted features of the model we employ
here are summarized as follows.

{\it The locus of the spiral arms.} The shape of the spiral arms is
one of the few characteristics upon which almost all observations
agree. In general, the spiral arms seem to be approximately
logarithmic; for the Milky Way, in particular, this seems to be the
case as shown in the very complete review by \citet{V05a}. We have
taken the spiral locus given by \citet{RHA79}, that combines a
straight bar in the center of the Galaxy and a smoothly joined
logarithmic region outwards.

{\it The number of spiral arms and the pitch angle.} We have taken the
bisymmetric fit of \citet{D00} to the K-band of the COBE satellite,
which has a pitch angle of $i\sim 15.5^o$. Although observations in
-H$\alpha$ \citep{RAA05}, pulsar rotation measures \citep{V05b}, radio
observations of HII regions (Sewilo et al. 2004; Paladini et
al. 2004), O and B stars (Negueruela \& Marco 2003), near IR flux and
gas \citep{BEG03}, and ultra-compact HII regions seem to show a
four-armed locus with a pitch angle $i\sim 13.5^o$ \citep{V05a},
we have assumed here that most of the mass is concentrated in the
stellar arms as seen in the K-band (which traces the more massive old
stellar population) rather than in the gaseous arms, which have been
proposed as being the response of the gas to this two-armed stellar
locus (Martos et al. 2004; Drimmel 2000). We expect that taking a
two-armed structure instead of a four-armed one will provide a good
(and conservative) estimate of the effects of this non-axisymmetric
structure on the globular cluster orbits.

{\it Outer limit for the spiral arms.} We have based the adopted
length of the spiral arms on observations rather than models. We have
taken for this parameter a limiting galactocentric distance of
$R_f\sim 12$ kpc (Drimmel 2000; Caswell \& Hanes 1987).

{\it Radial force produced by arms vs disk.} \citet{PC06} built a
family of self-consistent models and applied it to 12 normal spirals
with known rotation curves and photometry. They found a correlation
between the pitch angle and the ratio of the radial force of the arms
to that of the axisymmetric disk. For a galaxy like the Milky Way, the
pitch angles are in the range 11$^o$ to 16$^o$, from which they
obtained a force ratio of 4\% to 10\%.  This corresponds in our model
to a mass ratio of arms to disk of $M_A/M_D=[2\%,5\%]$. Independently,
models based on observations of our own Galaxy find a local force
ratio of approximately 4\% \citep{AL97}. In our model we adopt a local
force ratio of $\sim 5$\%, and a total mass for the bisymmetric spiral
arms of $M_A/M_D=0.03$.

{\it Angular velocity.} This is perhaps the most controversial
dynamical parameter, due to the intrinsic difficulty of measuring it.
Current observations and models, however, seem to show a trend,
placing its value in the interval $[20,25]$ km s$^{-1}$ kpc$^{-1}$.
Kinematic observations of stars of the local spiral arm, Orion, give
$\Omega = 25.7\pm 1.2$ km s$^{-1}$ kpc$^{-1}$ \citep{BGB06}, whereas
models (Martos et al. 2004; Pichardo et al. 2003) give $\Omega=20$ km
s$^{-1}$ kpc$^{-1}$. We have taken for this work $\Omega=20$ km
s$^{-1}$ kpc$^{-1}$.

\subsection{The Orbits Perturbed by Spiral Arms}\label{spiralresult}

Although individual spiral arms may be transient structures in spiral
galaxies, in fact the long-term effect of the spiral perturbation is
likely to be important, since even if they were to last only a short
time, they are soon replaced by new ones; this accounts for their
ubiquity in the majority of disk galaxies. It has been customary to
neglect the dynamical importance of spiral arms based on empirical or
intuitive arguments.  Here we perform computations to study
quantitatively their effects on the Galactic orbits, if any.

We integrated the six globular cluster orbits for the last 5 Gyr only,
because, as we shall show, the effects of the spiral perturbation are
already noticeable in this time span. The orbits are integrated in two
variants of the potential, both including the axisymmetric potential
(bulge, disk and halo). In one variant we added to the background
axisymmetric potential the spiral arms, and in the other, the spiral
arms and the bar.

\begin{figure}
\plotone{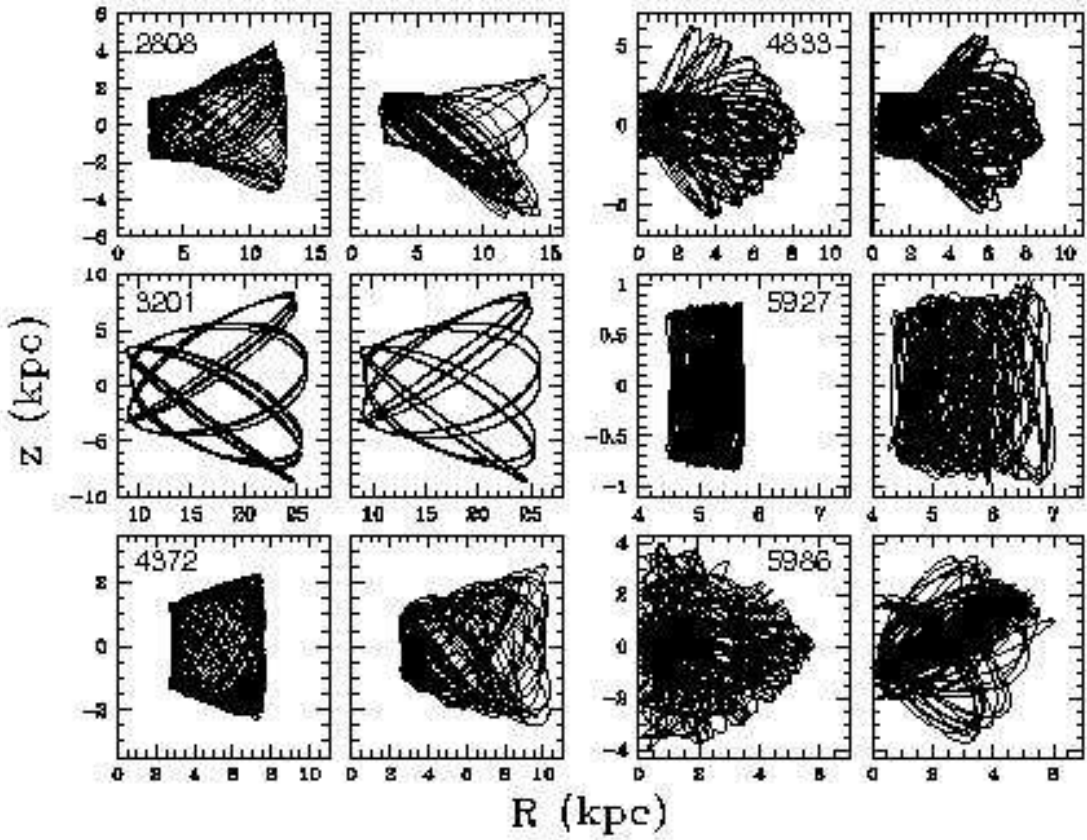}
\caption {The effect of the spiral arms. Meridional Galactic orbits of
the six clusters during the last 5 Gyr. For each cluster, the left
frame shows the orbit in the axisymmetric plus the two-spiral-arm
potential, and the right frame shows the orbit in the barred
(axisymmetric + bar potential) plus the two-spiral-arm potential.}
\label{mergalprb_arms}
\end{figure}

In Figure \ref{mergalprb_arms} we show the results of the computations
in the form of meridional orbits. The left panels refer to the orbits
computed with the first variant of the potential, i.e.  axisymmetric
plus two-spiral-arm potential, and the right panels to the orbits in
the second variant, i.e., axisymmetric plus bar plus two spiral arms.
In all cases we analyze the orbits integrated backwards in time.

If we compare directly the orbits in the pure axisymmetric potential
(Figure \ref{mergalorb_sh}) with those in the potential with spiral
arms (Figure \ref{mergalprb_arms}), we can clearly see the effect of
the spiral perturbation. The orbits change their shape in all cases,
although only for clusters remaining closest to the Galactic plane
(such as NGC 5927, NGC 4372, NGC 2808) the effect is noticeable.
Again, the least affected orbits are those that reach large radial and
vertical distances and remain mostly far from the influence of the
arms, as is the case of NGC 3201.  The orbit of this cluster is
practically the same in all model potentials.

The effect of the spiral perturbation is most pronounced in the case
of near-resonant (like that of NGC 2808), or irregular orbits (like
NGC 4833 and NGC 5986).  Indeed, NGC 2808 seems to get scattered out
of its present near-resonance orbit more readily by the spiral
perturbation than by the bar, a surprising result, since the mass of
the bar is much larger.  Another unexpected result is seen in the
orbit of NGC 4833, which seems more symmetric with respect to the
Galactic plane in the presence of the spiral perturbation.

As far as can be asserted on the basis of the six orbits here studied,
the orbital parameters, such as the maximum absolute values of the
$z$-distances and the peri- and apogalactic distances are not much
affected by the presence of the spiral arms.  The general tendency
seems to be for the cluster orbits to have larger energies in the
presence of spiral arms than in the axisymmetric case, and hence to
attain larger apogalactic distances.  This effect is largest for
clusters remaining close to the Galactic plane ($z\la 2$ kpc), and
inside the radial zone of strong influence of the spiral arms ($3<R<8$
kpc). The combined effects of bar and spiral arm perturbations seem to
be an efficient mechanism for attaining apogalactic distances larger
than in the axisymmetric potential.

Even though in general the dynamical effects of the bar dominate over
those of the spiral arms, it is interesting to see that the spiral
perturbations may play a non-negligible role. The clearest example is
that of NGC 2808.

\section{TIDAL RADII AND DESTRUCTION RATES}\label{tirad}
Ostriker et al. (1989) have emphasized the role played by a Galactic
bar in selectively destroying clusters in box orbits passing near the
Galactic center.  In our present study, as well as in Paper I, we find
that the smallest pericenters in the axisymmetric case (e.g. those of
NGC 4833 and NGC 5986) tend to be larger in the presence of the
bar. This is reflected in the destruction rates, which for these two
clusters are sensibly smaller in the presence of the bar.  Some
possible reasons for this discrepancy are discussed in Paper I.

In the same manner as in Paper I, we have assessed the effects of the
bar on the internal dynamics of the clusters and
computed tidal radii and destruction rates.

For the six clusters we computed tidal radii using the \citet{K62}
formula, $r_K=[M_c/M_g(3+e)]^{1/3}r_{min}$ with $M_c$ the mass of the
cluster, $M_g$ an ``effective Galactic mass'', $e$ the orbital
eccentricity, and $r_{min}$ the galactocentric distance of a
perigalactic point in the orbit. We have also calculated the tidal
radius of each cluster using an alternative approximate formula
introduced in Paper I, that has the advantage of not requiring an
assumption about an ``effective Galactic mass'', which is difficult to
define if the perigalactic point lies in a region with a strong
non-axisymmetric potential, as in the case of a bar. Our approximation
is given by the equation

\begin{equation}
  r_*=\left[{GM_c\over (\partial F_{x'}/\partial x')_{r'=0}+\dot\theta^2+\dot\varphi^2\sin^2\theta}\right]^{1/3},
\label{redm}
\end{equation}

\noindent where $F_{x'}$ is the component of the Galactic acceleration
along the line $x'$ that joins the cluster with the Galactic center,
and its partial derivative is evaluated at the position of the
cluster. The angles $\varphi$ and $\theta$ are angular spherical
coordinates of the cluster in an inertial Galactic frame. The tidal
radius is computed at a perigalactic point in the orbit. In a two body
problem $r_*$ is reduced to King's formula. For the derivation of the
equation \ref{redm} see Appendix A of Paper I.

The last three columns of Table \ref{tbl-1} contain $r_K$, $r_*$, and
$r_L$ for the axisymmetric potential; $r_L$ is the observed limiting
radius of the cluster. In the last two columns of Table \ref{tbl-2},
we give $r_K$ y $r_*$ for the barred potential.

\begin{figure}
\plotone{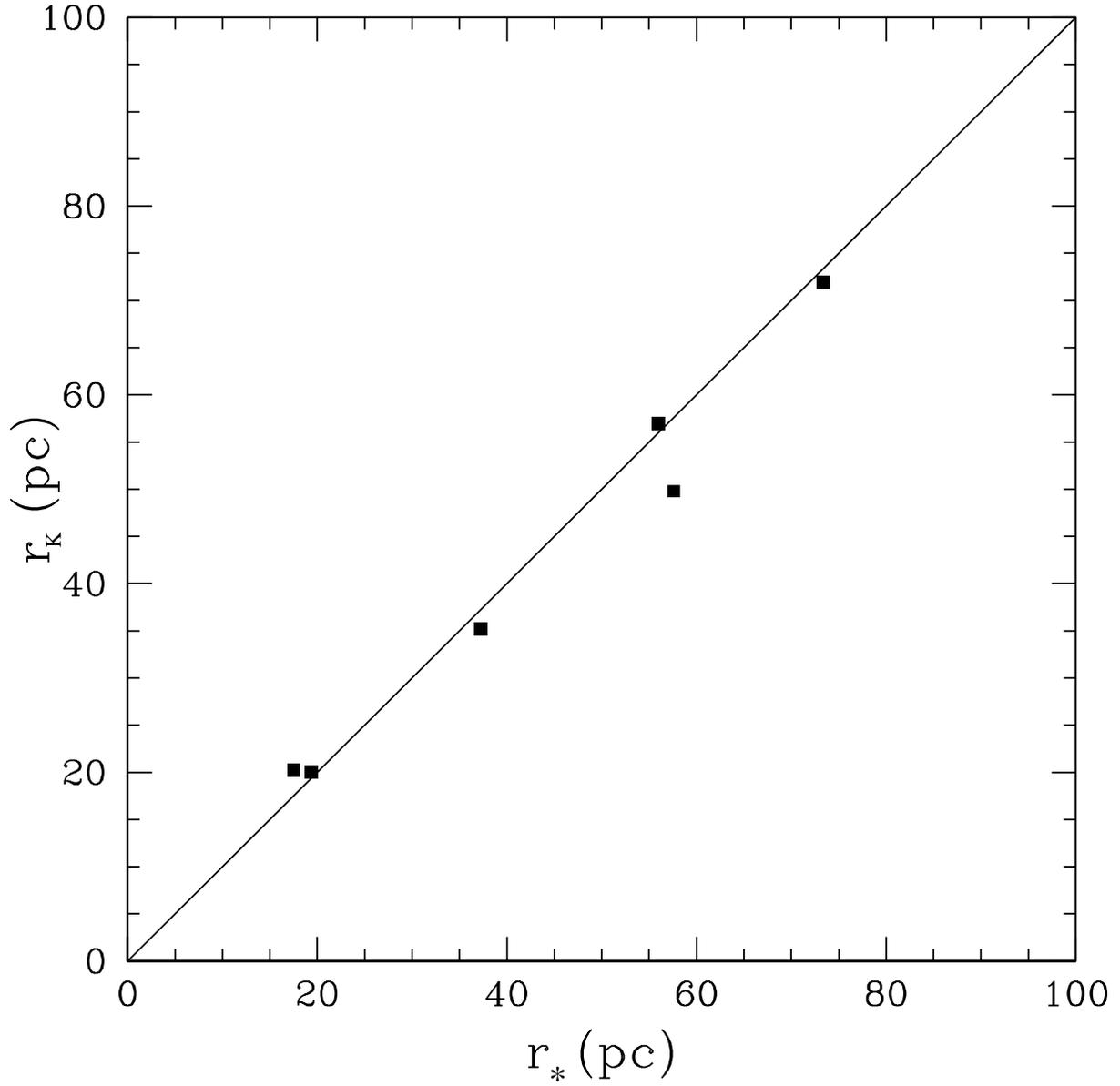}
\caption {Comparison of tidal radii $r_k$ and $r_*$ using the
  non-axisymmetric (barred) potential. The line is the line of
  coincidence.}
\label{rkrs_bar}
\end{figure}

In Paper I we found that the formulae for $r_*$ and $r_K$ gave similar
values.  The same is true for this new set of objects, independently
of the potential variant employed (barred, barred plus spiral, or
axisymmetric).  This is shown in Figure \ref{rkrs_bar}, where we have
taken the barred model as an example. In Figure \ref{rkrs_bar} we show
a comparison of the observed limiting radii ($r_L$) with the
theoretical ones; here using in particular King's formula ($r_K$).

\begin{figure}
\plotone{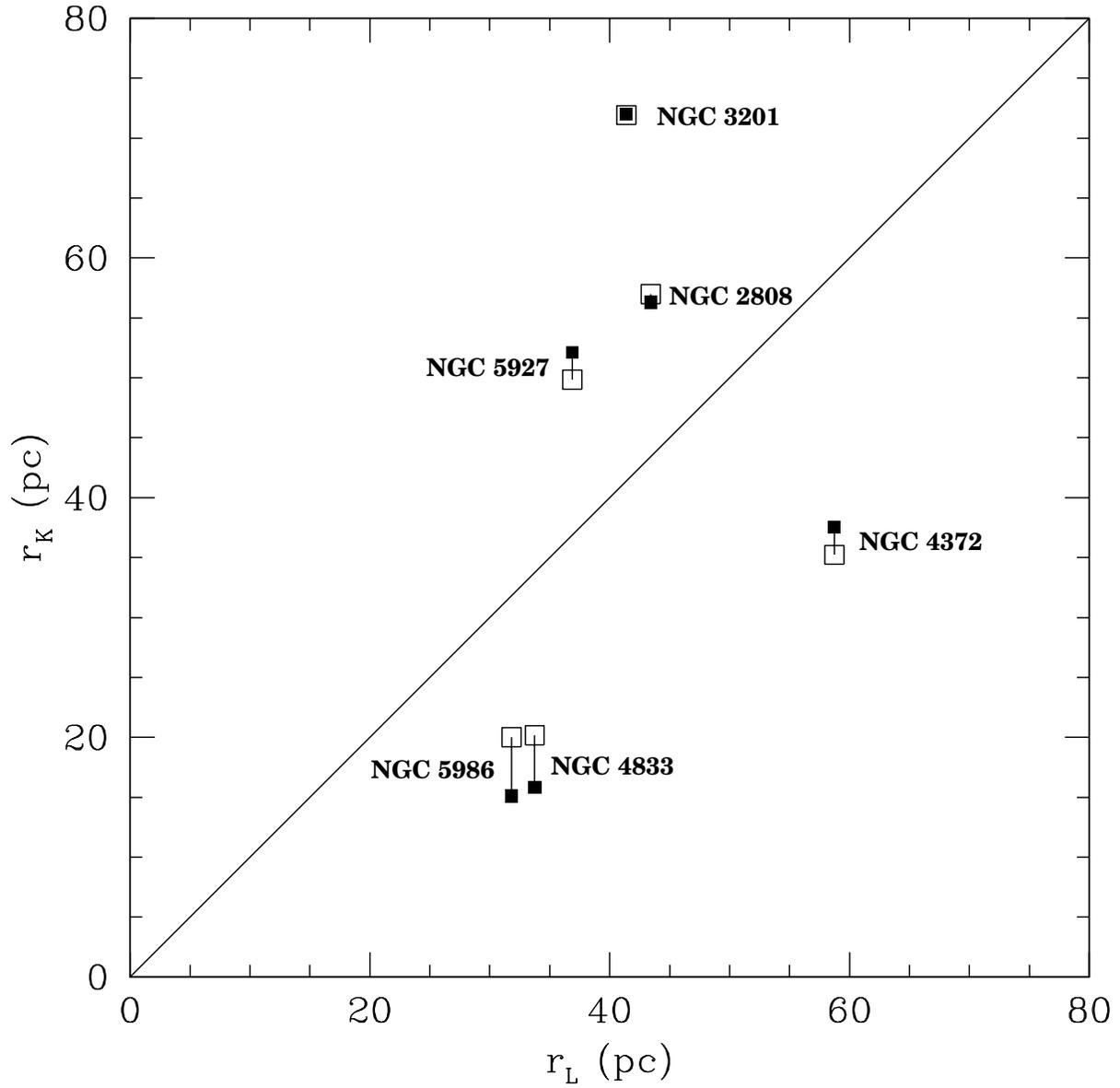}
\caption {Tidal radius $r_{K}$ (King) compared with the observed
limiting radius $r_{L}$. Results with the axisymmetric (filled
squares), and barred (empty squares) potentials. The line is the line
of coincidence.}
\label{rkrl_bar}
\end{figure}

The prograde-retrograde cluster NGC 5986 has a mean perigalactic
distance that places it within the bar. Its tidal radius in Figure
\ref{rkrl_bar} lies under the line of coincidence, as was the case for
most of the retrograde or prograde-retrograde clusters with a mean
perigalactic distance of less than 3.5 kpc studied in Paper I.  The
only fully retrograde globular cluster of the new sample, NGC 3201,
lies well above the line of coincidence but it is also the only
cluster residing well outside the bar. For a discussion of the
importance of the orbital sense (prograde or retrograde) on the tidal
radii we refer the reader to Section 4 of Paper I.

Regarding the destruction rates, we have calculated the time of
destruction due to components of bulge and disk in the Galaxy based on
the results of several papers on the subject (Aguilar et al. 1988;
Long et al. 1992; Gnedin \& Ostriker 1997,1999; Spitzer 1987; Kundi\'c
\& Ostriker 1995; Gnedin et al. 1999a,b). As in Paper I, we define the
total destruction rate due to gravitational shocks with the bulge as

\begin{equation} \frac{1}{t_{bulge}} = \frac{1}{t_{bulge,1}} +
\frac{1}{t_{bulge,2}}
\label{tbtot} 
\end{equation}
 
\noindent where

\begin{equation} t_{bulge,1} = \left ( \frac{-E_c}{<\!(\Delta E)_b\!>}
\right )P_{orb}, \label{tb1} \end{equation}

\begin{equation} t_{bulge,2} = \left ( \frac{E_c^2}{<\!(\Delta E)_b^2\!>}
\right )P_{orb}, \label{tb2} \end{equation}

\noindent with $E_c \simeq -0.2GM_c/r'_h$ the mean binding energy per
unit mass of the cluster and $P_{orb}$ its orbital period in the
Galactic radial direction.

In the same manner, for the disk we have the total destruction
rate due to gravitational shocks,

\begin{equation} \frac{1}{t_{disk}} = \frac{1}{t_{disk,1}} +
\frac{1}{t_{disk,2}}, \label{tdtot} \end{equation}

\noindent where

\begin{equation} t_{disk,1} = \left ( \frac{-E_c}{<\!(\Delta E)_d\!>}
\right ) \frac{P_{orb}}{n}, \label{td1} \end{equation}

\begin{equation} t_{disk,2} = \left ( \frac{E_c^2}{<\!(\Delta E)_d^2\!>}
\right ) \frac{P_{orb}}{n}. \label{td2} \end{equation}

\noindent Here, $n$ is the number of crossings with the disk during
the radial orbital period $P_{orb}$. For a summary on destruction rate
formulae see Paper I.

We have applied these formulae to the six globular clusters and to the
specific case including a Galactic bar. It is worth noting that we
have taken very conservative values for the mass and radius for the
Galactic bar.  According to recent estimates the bar could extend at
least 1 kpc beyond the limit we used (L\'opez-Corredoira et al.
2007), and it could have at least twice the observed mass if it is
combined with the response of a dark matter halo, which apparently
produces a dark bar aligned with the visible one (Col{\'{\i}}n et al.
2006; Athanassoula 2007); thus the effect of the bar might be
considerably stronger.

In Table \ref{tbl-3} we present the destruction rates obtained in our
computations. The first column shows the name of the cluster. The
second column gives the mass of the cluster, computed with a
mass-to-light ratio of 2. Absolute visual magnitudes are taken from
\citet{H96}. The central concentration and the half-mass
radius are given in columns three and four, and are taken from
\citet{H96}.

The remaining columns show the destruction rates due to the bulge and
disk. These destruction rates
are the averages over the last $10^9$ yr in a cluster's orbit, taking
the corresponding perigalactic points and crossings with the Galactic
disk. For a given cluster, the first line in Table \ref{tbl-3} shows
the destruction rates computed in the axisymmetric potential; the
second line gives the corresponding values in the barred potential.

The new sample resides much closer to the Galactic disk than did most
of the 48 clusters of Paper I.  NGC 3201 and NGC 2808 show the longest
total destruction times, taking both disk and bulge into account. The
difference between the barred and non-barred models is largest for NGC
5986, NGC 4833, whose orbits lie almost entirely within the bar, and
are thus the most affected by it.  Also affected though not as much,
are NGC 5927 that resides in the outskirts of the bar but very close
to the Galactic plane, and NGC 4372 whose average apocenter lies
inside the bar.

\begin{figure}
\plotone{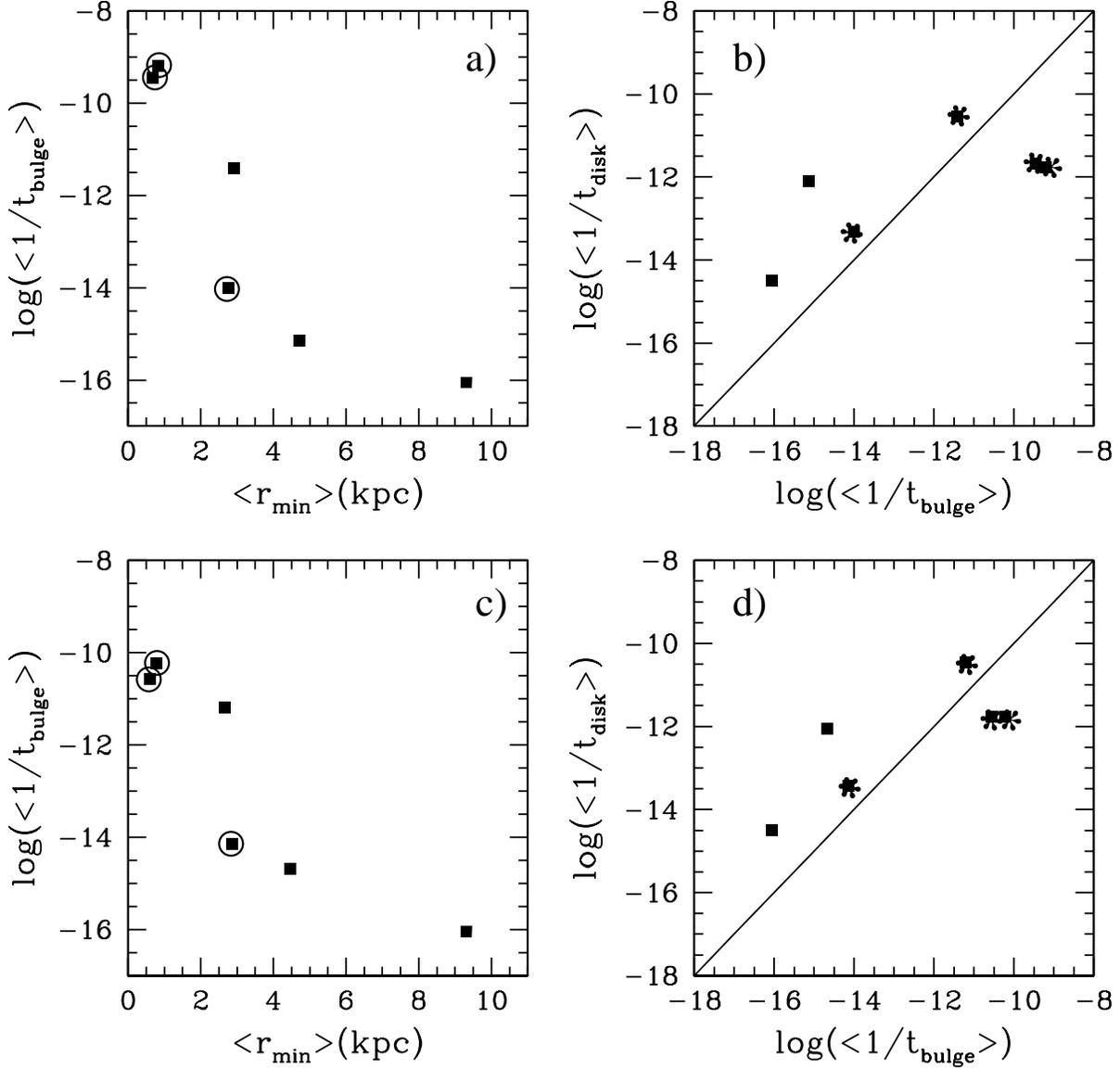}
\caption {Destruction rates with the central orbits.  We show the
values of the total destruction rate due to the bulge as a function of
$<$$r_{min}$$>$, the average minimum distance to the Galactic center,
in (a) the axisymmetric, and (c) the non-axisymmetric potential. The
marked squares correspond to clusters with $<$$r_{min}$$>$ $<$ 3 kpc
and orbital eccentricity $e > 0.6$.  In (b) and (d) we give the
comparison of the destruction rates due to the bulge and disk, in the
axisymmetric and non-axisymmetric potentials, respectively. Marked
squares correspond to clusters with $<$$r_{min}$$>$ $<$ 3 kpc; the
line of coincidence is plotted in both panels.}
\label{fdrates}
\end{figure}

In Figure \ref{fdrates} we plot only the values obtained with the
`central orbits'. This figure shows that the destruction rates due to
the bulge depend on the mean perigalactic distance, both for the
axisymmetric potential (Frame a) and for the barred potential (Frame
c). Clusters with perigalactic points close to the Galactic center and
large orbital eccentricities (squares marked with a circle) have, in
general, greater bulge destruction rates. Frames (b) and (d) of Figure
\ref{fdrates} show the comparison of the destruction rates due to the
bulge and disk, in the axisymmetric and in the barred potential,
respectively. These frames show that bulge shocking dominates in the
bar region, as found by \citet{AHO88}. We confirm as in Paper I that
the destruction rates are very similar in both the axisymmetric and
the barred Galactic potential.

In this study we have found that the bar is the most important
non-axisymmetric structure in the Galaxy as regards its orbital
effects, tidal radii and destruction rates of globular clusters. The
spiral arms produce small, though non-negligible effects in the orbits
of the clusters moving closest to the Galactic plane (see Section
\ref{spiralresult}). Since the changes in destruction rates and tidal
radii depend mainly on the orbital parameters, the effects of the
spiral arms in a Milky-Way-like galaxy are negligible compared to the
effects of the bar. However, in the case of grand-design spirals or of
galaxies with a weak bar or entirely without bar, the effects of the
spiral arms should not be considered to be negligible.

\section{DISCUSSION AND CONCLUSIONS}\label{conclusiones}

We have obtained orbits for 6 additional globular clusters in both a
barred and an axisymmetric Galactic potential, as well as in models
including spiral arms. The total number of clusters with available
orbits is now 54. Among the newly calculated cases, only the orbit of
NGC 3201, an ``outer'' cluster, is unaffected by the bar. The other
five clusters have orbits residing within the region of influence of
the bar, and their orbits are clearly influenced by it. As in our
previous study, we find that the main changes the bar causes in the
orbits are larger vertical and radial excursions, and far more
irregular orbits. In general, the bar causes no net global changes in
the energy or the $z$-component of the angular momentum, computed in
an inertial frame.  However, in four out of the six cases, jumps in
these quantities do occur, even causing a temporary reversal of the
sense of rotation of the orbit in the case of NGC 5986. The effect of
a shorter-lived bar is found to be quite as noticeable on these orbits
as that of the longer-lived one.

Orbits with spiral-arm perturbations were also computed. Contrary to
expectations, even a small perturbation, accounting for only 5\% of
the local radial force ratio arm-disk, causes significant changes in
the form of the orbits. Although the influence of the spiral arms on
the orbits of the clusters closest to the Galactic disk are not
negligible, the long-term effects are definitely dominated by the
Galactic bar.

Tidal radii have been computed with the expression derived in Paper I,
as well as with a numerical evaluation of the relevant quantities
along the orbit. Again, we found little change due to the bar. When
changes did occur, they again make the computed tidal radii somewhat
larger in the presence of a bar. With the new material on hand, we
confirm our earlier finding that the destruction rates due to shocks
with the Galactic bulge and disk are not strongly affected when we
consider a barred Galactic potential, or a barred potential with
spiral arm perturbations.

\section*{Acknowledgments}
B.P. thankfully acknowledges the support of CONACyT, Mexico, through
grant 50720.

\clearpage
\begin{deluxetable}{ccccccccccccc}
\tabletypesize{\scriptsize}
\rotate
\tablecaption{Orbits with the axisymmetric potential \label{tbl-1}}
\tablewidth{0pt}
\tablehead{
\colhead{Cluster} & \colhead{$(r_{min})_{min}$} & \colhead{$<$$r_{min}$$>$} &
\colhead{$(r_{max})_{max}$} & \colhead{$<$$r_{max}$$>$} & \colhead{($|z|_{max})
_{max}$} & \colhead{$<$$|z|_{max}$$>$} & \colhead{$<$$e$$>$} & \colhead{$E$} &
\colhead{$h$} & \colhead{$r_{K}$} & \colhead{$r_{\ast}$} &
\colhead{$r_{L}$} \\ \colhead{} & \colhead{($kpc$)} & \colhead{($kpc$)} &
\colhead{($kpc$)} & \colhead{($kpc$)} & \colhead{($kpc$)} &
\colhead{($kpc$)} & \colhead{} & \colhead{(10$kms^{-1})^2$} &
\colhead{(10$kms^{-1}kpc$)} & \colhead{($pc$)} & \colhead{($pc$)} &
\colhead{($pc$)}
}
\startdata
NGC 2808 & 2.65 & 2.76 & 12.86 & 12.72 & 4.42 & 2.60 & 0.643 & $-1192.93$ &
87.32 & 56.3 & 54.3 & 43.4 \\
         & 1.88 & 2.10 & 11.82 & 11.48 & 4.86 & 2.38 & 0.691 & $-1241.68$ &
68.36 & 48.7 & 46.1 & \\
         & 3.67 & 3.89 & 13.83 & 13.81 & 5.89 & 3.95 & 0.560 & $-1134.84$ &
112.24 & 72.7 & 73.0 & \\
NGC 3201 & 9.26 & 9.31 & 27.02 & 26.99 & 8.97 & 5.95 & 0.487 & $-841.75$ &
$-278.09$ & 72.0 & 73.4 & 41.4 \\
         & 9.03 & 9.08 & 22.99 & 22.96 & 7.15 & 4.99 & 0.433 & $-895.62$ &
$-262.22$ & 70.6 & 73.4 & \\
         & 9.50 & 9.56 & 31.99 & 31.97 & 11.24 & 7.15 & 0.540 & $-784.32$ &
$-295.27$ & 73.0 & 73.4 & \\
NGC 4372 & 2.91 & 2.93 & 7.88 & 7.74 & 2.06 & 1.74 & 0.451 & $-1377.43$ &
83.03 & 37.5 & 38.6 & 58.8 \\
         & 1.99 & 2.04 & 7.72 & 7.62 & 1.64 & 1.35 & 0.578 & $-1416.52$ &
64.81 & 29.0 & 28.3 & \\
         & 4.06 & 4.22 & 7.74 & 7.74 & 2.89 & 2.37 & 0.294 & $-1330.13$ &
102.25 & 48.2 & 52.8 & \\
NGC 4833 & 0.19 & 0.68 & 8.92 & 8.08 & 6.02 & 1.80 & 0.847 & $-1410.26$ &
11.04 & 15.8 & 14.7 & 33.8 \\
         & 0.01 & 0.76 & 8.13 & 6.71 & 6.31 & 3.32 & 0.808 & $-1453.00$ &
0.33 & 13.1 & 11.3 & \\
         & 0.48 & 0.96 & 9.80 & 9.06 & 5.82 & 2.05 & 0.811 & $-1355.97$ &
25.27 & 18.4 & 17.1 & \\
NGC 5927 & 4.70 & 4.72 & 5.76 & 5.74 & 0.84 & 0.80 & 0.098 & $-1422.14$ &
107.14 & 52.1 & 61.5 & 36.9 \\
         & 3.85 & 3.85 & 5.46 & 5.46 & 0.67 & 0.62 & 0.172 & $-1479.97$ &
93.17 & 45.5 & 52.9 & \\
         & 4.68 & 4.71 & 7.41 & 7.39 & 1.23 & 1.08 & 0.222 & $-1344.77$ &
119.58 & 51.3 & 58.2 & \\
NGC 5986 & 0.08 & 0.82 & 5.67 & 4.44 & 3.97 & 1.98 & 0.694 & $-1622.48$ &
4.04 & 15.1 & 14.5 & 31.8 \\
         & 0.00 & 0.67 & 4.98 & 3.83 & 3.61 & 1.92 & 0.714 & $-1681.18$ &
0.00 & 16.5 & 16.1 & \\
         & 0.99 & 1.07 & 5.93 & 5.79 & 2.88 & 1.74 & 0.688 & $-1545.02$ &
17.96 & 21.0 & 20.8 & \\
\enddata
\end{deluxetable}

\clearpage
\begin{deluxetable}{cccccccccccc}
\tabletypesize{\scriptsize}
\rotate
\tablecaption{Orbits with the barred potential \label{tbl-2}}
\tablewidth{0pt}
\tablehead{
\colhead{Cluster} & \colhead{$(r_{min})_{min}$} & \colhead{$<$$r_{min}$$>$} &
\colhead{$(r_{max})_{max}$} & \colhead{$<$$r_{max}$$>$} & \colhead{$(|z|_{max})
_{max}$} & \colhead{$<$$|z|_{max}$$>$} & \colhead{$<$$e$$>$} & \colhead{$h_{min}$}
 & \colhead{$h_{max}$} & \colhead{$r_{K}$} & \colhead{$r_{\ast}$} \\
\colhead{} & \colhead{($kpc$)} & \colhead{($kpc$)} &
\colhead{($kpc$)} & \colhead{($kpc$)} & \colhead{($kpc$)} &
\colhead{($kpc$)} & \colhead{} & \colhead{(10$kms^{-1}kpc$)} &
\colhead{(10$kms^{-1}kpc$)} & \colhead{($pc$)} & \colhead{($pc$)}
}
\startdata
NGC 2808 & 2.37 & 2.86 & 15.63 & 13.27 & 5.67 & 2.92 & 0.645 & 77.67 &
101.11 & 57.0 & 56.0 \\
         & 1.40 & 1.82 & 12.83 & 11.33 & 5.17 & 1.54 & 0.723 & 55.01 &
79.13 & 42.5 & 39.9  \\
         & 3.35 & 3.93 & 14.31 & 11.99 & 6.21 & 3.97 & 0.505 & 93.15 &
114.73 & 72.4 & 73.0  \\
NGC 3201 & 9.26 & 9.31 & 26.99 & 26.96 & 8.96 & 5.94 & 0.487 & $-278.21$ &
$-278.03$ & 71.9 & 73.4  \\
         & 9.03 & 9.07 & 22.97 & 22.94 & 7.14 & 4.98 & 0.433 & $-262.34$ &
$-262.15$ & 70.6 & 73.3  \\
         & 9.50 & 9.56 & 31.97 & 31.94 & 11.23 & 7.15 & 0.539 & $-295.39$ &
$-295.22$ & 72.9 & 73.4  \\
NGC 4372 & 1.91 & 2.65 & 8.03 & 6.60 & 3.04 & 1.59 & 0.428 & 47.37 &
88.84 & 35.2 & 37.2  \\
         & 1.41 & 1.78 & 7.67 & 6.91 & 3.23 & 1.09 & 0.590 & 47.27 &
69.36 & 28.0 & 27.8  \\
         & 3.48 & 3.92 & 9.61 & 7.64 & 3.36 & 2.09 & 0.317 & 85.32 &
112.12 & 43.7 & 46.7  \\
NGC 4833 & 0.19 & 0.79 & 10.56 & 8.66 & 5.85 & 1.78 & 0.835 & 0.96 &
30.57 & 20.2 & 17.5  \\
         & 0.14 & 0.81 & 11.85 & 8.83 & 5.96 & 2.04 & 0.833 & $-4.71$ &
37.21 & 17.6 & 14.5  \\
         & 0.36 & 1.19 & 11.25 & 9.28 & 6.35 & 2.60 & 0.777 & 10.32 &
41.41 & 20.8 & 18.6  \\
NGC 5927 & 4.28 & 4.47 & 6.64 & 5.98 & 0.90 & 0.79 & 0.144 & 103.19 &
112.55 & 49.8 & 57.6  \\
         & 0.83 & 2.64 & 7.26 & 4.32 & 0.63 & 0.45 & 0.252 & 23.67 &
108.04 & 28.3 & 32.4  \\
         & 4.67 & 4.70 & 7.67 & 7.50 & 1.25 & 1.09 & 0.229 & 117.90 &
121.30 & 51.5 & 58.5  \\
NGC 5986 & 0.03 & 0.60 & 6.50 & 5.21 & 3.53 & 1.23 & 0.795 & $-6.80$ &
19.94 & 20.0 & 19.4 \\
         & 0.02 & 0.58 & 5.60 & 4.45 & 3.03 & 1.05 & 0.775 & $-20.88$ &
16.90 & 19.4 & 18.8  \\
         & 0.10 & 0.90 & 6.69 & 5.10 & 4.14 & 1.90 & 0.708 & $-9.40$ &
26.12 & 23.3 & 22.7  \\
\enddata
\end{deluxetable}

\clearpage
\begin{deluxetable}{cccccccccc}
\tabletypesize{\scriptsize}
\rotate
\tablecaption{Destruction Rates \label{tbl-3}}
\tablewidth{0pt}
\tablehead{
\colhead{Cluster} & \colhead{$M_c$} & \colhead{$c$} & \colhead{$r_h$} &
\colhead{$<$$1/t_{bulge,1}$$>$} &
\colhead{$<$$1/t_{bulge,2}$$>$} & \colhead{$<$$1/t_{bulge}$$>$} &
\colhead{$<$$1/t_{disk,1}$$>$} & \colhead{$<$$1/t_{disk,2}$$>$} &
\colhead{$<$$1/t_{disk}$$>$} \\ \colhead{} & \colhead{($M_{\odot}$)} &
\colhead{} & \colhead{($pc$)} &
\colhead{($yr^{-1}$)} & \colhead{($yr^{-1}$)} & \colhead{($yr^{-1}$)} & 
\colhead{($yr^{-1}$)} & \colhead{($yr^{-1}$)} & \colhead{($yr^{-1}$)}
}
\startdata
NGC 2808 & 9.7E05 & 1.77 & 2.12 & 9.5E-15 & 6.3E-16 & 1.0E-14 & 4.4E-14 &
3.5E-15 & 4.7E-14 \\
 &  &  &  & 6.9E-15 & 4.5E-16 & 7.3E-15 & 3.3E-14 & 2.7E-15 & 3.6E-14 \\
NGC 3201 & 1.6E05 & 1.30 & 3.90 & 8.1E-17 & 8.1E-18 & 8.9E-17 & 2.8E-15 &
3.9E-16 & 3.2E-15 \\
 &  &  &  & 8.1E-17 & 8.2E-18 & 9.0E-17 & 2.8E-15 & 3.9E-16 & 3.2E-15 \\
NGC 4372 & 2.2E05 & 1.30 & 6.58 & 3.1E-12 & 7.2E-13 & 3.9E-12 & 2.2E-11 &
6.3E-12 & 2.9E-11 \\
 &  &  &  & 5.1E-12 & 1.2E-12 & 6.3E-12 & 2.7E-11 & 7.8E-12 & 3.5E-11 \\
NGC 4833 & 3.1E05 & 1.25 & 4.56 & 2.5E-10 & 1.0E-10 & 3.5E-10 & 1.6E-12 &
4.8E-13 & 2.1E-12 \\
 &  &  &  & 4.3E-11 & 1.7E-11 & 6.0E-11 & 1.3E-12 & 4.2E-13 & 1.7E-12 \\
NGC 5927 & 2.2E05 & 1.60 & 2.54 & 6.8E-16 & 5.0E-17 & 7.3E-16 & 7.2E-13 &
7.3E-14 & 8.0E-13 \\ 
 &  &  &  & 1.9E-15 & 1.5E-16 & 2.1E-15 & 8.0E-13 & 8.2E-14 & 8.8E-13 \\
NGC 5986 & 4.1E05 & 1.22 & 3.18 & 4.8E-10 & 1.6E-10 & 6.4E-10 & 1.4E-12 &
3.0E-13 & 1.7E-12 \\
 &  &  &  & 2.2E-11 & 5.0E-12 & 2.7E-11 & 1.4E-12 & 2.7E-13 & 1.7E-12 \\
\enddata
\end{deluxetable}

\clearpage

\clearpage

\clearpage

\clearpage

\clearpage

\clearpage

\clearpage

\clearpage

\clearpage

\clearpage

\clearpage

\end{document}